\documentclass[useAMS,usenatbib]{mn2e}
\usepackage{graphicx, setspace, subfigure, latexsym, amssymb, amsmath, booktabs, paralist}
\usepackage{gensymb}

\title[A 6.5-GHz Multibeam Pulsar Survey]{A 6.5-GHz Multibeam Pulsar Survey}
\author[S. D. Bates et al.]
	{S. D. Bates$^{1}$\thanks{Email: samuel.d.bates@student.manchester.ac.uk}, S. Johnston$^{2}$, D.  R. Lorimer$^{3,4}$, M. Kramer$^{1,5}$, A. Possenti$^{6}$, \newauthor M. Burgay$^{6}$, B. Stappers$^{1}$, M. J. Keith$^{2}$, A. Lyne$^{1}$, M. Bailes$^{7}$, M. A. McLaughlin$^{3,4}$, \newauthor J. T. O'Brien$^{1}$ and G. Hobbs$^{2}$ \\
$^{1}$Jodrell Bank Centre for Astrophysics, School of Physics and Astronomy, The University of Manchester, Manchester M13 9PL, UK\\
$^{2}$Australia Telescope Nation Facility, CSIRO, P.O. Box 76, Epping NSW 1710, Australia\\
$^{3}$Department of Physics, West Virginia University, 210 Hodges Hall, Morgantown, 26506 USA\\
$^{4}$National Radio Astronomy Observatory, PO Box 2, Green Bank, WV 24944, USA\\
$^{5}$MPI fuer Radioastronomie, Auf dem Huegel 69, 53121 Bonn, Germany\\
$^{6}$INAF - Osservatorio Astronomico di Cagliari, Poggio dei Pini, 09012 Caopterra, Italy\\
$^{7}$Centre for Astrophysics and Supercomputing, Swinburne University of Technology, PO Box 218 Hawthorn, VIC 3122, Australia}
\begin{document}

\date{Accepted: 29 September 2010}

\pagerange{\pageref{firstpage}--\pageref{lastpage}} \pubyear{2010}

\maketitle

\label{firstpage}
\begin{abstract}
A survey of the Galactic plane in the region $-60\degree \leq l \leq 30\degree$, $|b| \leq 0.25\degree$ was carried out using the seven-beam Parkes Methanol Multibeam (MMB) receiver, which operates at a frequency of 6.5~GHz.
Three pulsars were discovered, and 16 previously known pulsars detected. In this paper we present two previously-unpublished discoveries, both with extremely high dispersion measures, one of which is very close, in angular distance, to the Galactic centre. The survey data also contain the first known detection, at radio frequencies, of the radio magnetar PSR J1550$-$5418. Our survey observation was made 46 days prior to that previously published and places constraints on the beginning of pulsed radio emission from the source.
%Three pulsars were discovered, and 16 previously known pulsars detected. One of the newly-discovered pulsars, associated with EGRET source 3EG J1410$-$6147, was found in the first pass through the survey data. In this paper we present two previously-unpublished discoveries, both with extremely high dispersion measures, one of which is very close, in angular distance, to the Galactic centre. The survey data also contain the first known detection, at radio frequencies, of the radio magnetar PSR J1550$-$5418. Our survey observation was made 46 days prior to that previously published and places constraints on the beginning of pulsed radio emission from the source.

The detection of only three previously undiscovered pulsars argues that there are few pulsars in the direction of the inner Galaxy whose flux density spectrum is governed by a flat power law. However, these pulsars would be likely to remain undetected at lower frequencies due to the large amount of scatter broadening which affects pulsars with high values of dispersion measure. Surveys with future telescopes at high observing frequencies will, therefore, play an important role in the discovery of pulsars at the Galactic centre. By simulating pulsar surveys of the Galaxy with Phase 1 SKA at frequencies of 1.4~GHz and 10~GHz, we find that high-frequency observations are the only way to discover and observe the Galactic-centre pulsar population.

\end{abstract}


\section{Introduction}
Pulsars are known to have steep spectra in the radio regime. At frequencies above about $100\,\mathrm{MHz}$, the relationship between flux density, $S_\nu$, and observing frequency, $\nu$, can be approximated by the power law $S_\nu \propto \nu^{\alpha}$ with a spectral index $\alpha$. From a large sample of pulsars, \citet{maron2000} derived a mean value for $\alpha$ of $-1.8$. For this reason, pulsar surveys have frequently been conducted at frequencies $\lesssim$1500~MHz - for example the Parkes southern pulsar survey \citep{mld+95} at 436~MHz and the Green Bank pulsar survey at 350~MHz \citep{hessels2008}, while the higher frequency of 1400~MHz was used in the Swinburne intermediate-latitude pulsar survey \citep{ebsb01}, the Parkes multibeam pulsar survey \citep[PMPS, ][]{mlc+01} and the PALFA survey at Arecibo \citep{cordes2006}. In addition, the beamwidth of a telescope scales as $\nu^{-1}$ and so the amount of time required to survey a given area of sky with a fixed time per pointing is reduced.

Charged particles in the interstellar medium cause a signal of frequency $f$ to be delayed, relative to a signal at infinite frequency, by a time
\begin{equation}
t=\frac{e^2}{2\pi m_e c}\frac{\rm{DM}}{f^2},
\label{timedelay}
\end{equation}
where we define the dispersion measure, DM, to be
\begin{equation}
DM=\int_0^d n_\mathrm{e} \; \rm{d}l
\label{defineDM}
\end{equation}
for an electron number density $n_\mathrm{e}$ and path length from the pulsar to Earth, $d$. At low frequencies, therefore, the dispersive delay is large, making it easier to discern genuine astrophysical signals at low DMs from terrestrial radio frequency interference (RFI).

However, at low frequencies, there are two factors which can have a strong limiting effect on the sensitivity of a survey. The first is the $\nu^{-2.6}$ dependence of the sky background temperature, T$_{\mathrm{sky}}$, \citep[e.g.\,][]{lmop87}, which arises due mainly to synchrotron radiation from free electrons in the Galactic magnetic field. This can dominate the system temperature, T$_{\mathrm{sys}}$, at low frequencies, reducing sensitivity to all pulsars - especially those at low Galactic latitudes.

The second of these factors is interstellar scattering, which causes a signal to be broadened with a typically exponential tail of timescale $\tau_s$, which scales with frequency approximately as $\tau_s \propto \nu^{-3.5}$ \citep{lkm+01,bcc+04}. Unlike dispersion, this scatter broadening cannot be easily removed in the signal processing. Therefore, at higher frequencies a survey of the Galactic plane will be less affected by smearing of pulses from distant sources - albeit at the cost of reduced intrinsic flux density, but mitigated by reduced sky background temperature and larger available bandwidth.

This effect is important, and provided two motivations for undertaking this high-frequency survey of the Galactic plane. Firstly, the distribution of pulsars in the inner Galaxy is not well known. Following earlier work by \citet{joh94}, \citet{lfl+06} showed that there may be a dearth of pulsars in this region, but that this result is highly dependent on the poorly-known free-electron distribution. Finding pulsars in this region, through surveys at higher frequencies \citep[e.g.\,][]{joh2006}, 
can constrain both the distribution of pulsars and dispersive material.

Secondly, it has long been known that many exciting tests of General Relativity, as well as a large amount of evolutionary information, would be facilitated by the discovery of a binary system containing both a pulsar and a black hole \citep{kramer2004}. By looking towards the Galactic centre, one might hope to observe a pulsar in a long orbit around the supermassive black hole, or compact stellar-mass binaries produced in the high stellar density conditions \citep{pfahl2004}.

In this paper we present the findings of the first wide-area pulsar survey at a frequency above 5~GHz. In Section 2 we discuss the parameters of the survey and the resulting sensitivity limit of our observations, and briefly mention the processing of the data. In Section 3 we explain the problems that arise from searching for pulsars at this high frequency and the techniques used to minimise these difficulties. In Section 4 we outline the parameters of the newly discovered pulsars, and then look at redetections of known pulsars in the survey area and use these to measure the variation of scattering timescale with dispersion measure (hereafter DM) and how the scattering and spectral index affect what pulsars we are able to observe. Finally, in Section 5, we summarize our main conclusions.

\begin{table}
	\begin{center}
	\caption{Observational parameters for the MMB pulsar survey.}
		\begin{tabular}{lr}
		\toprule
		%\multicolumn{2}{c}{Receiver Parameters} \\
		Number of beams  & 7 \\
		Polarizations/beam   & 2 \\
		Centre Frequency (MHz) & 6591\\
		Frequency channels  & 192 $\times$ 3 MHz \\
		System temperature, $T_{\rm sys}$ & 40 K \\
		Gain, G & 0.6 $\mathrm{K\,Jy^{-1}}$\\
		Half-power beamwidth (deg) & 0.053 \\
		\midrule
		%\multicolumn{2}{c}{Survey Parameters} \\
		Galactic longitude range & $-60\degree$ to $30\degree$ \\
		Galactic latitude range & $-0.25\degree$ to $0.25\degree$ \\
		Number of survey pointings & 2560\\
		Sampling interval & 125 $\mathrm{\mu}$s \\
		Observation time/pointing, $t_{\rm obs}$& 1055 s \\
		Limiting sensitivity & 0.15 mJy \\
		\bottomrule
		\end{tabular}
		\label{parameters}
	\end{center}
\end{table}

\begin{figure}
	\begin{center}	\includegraphics[width=8cm]{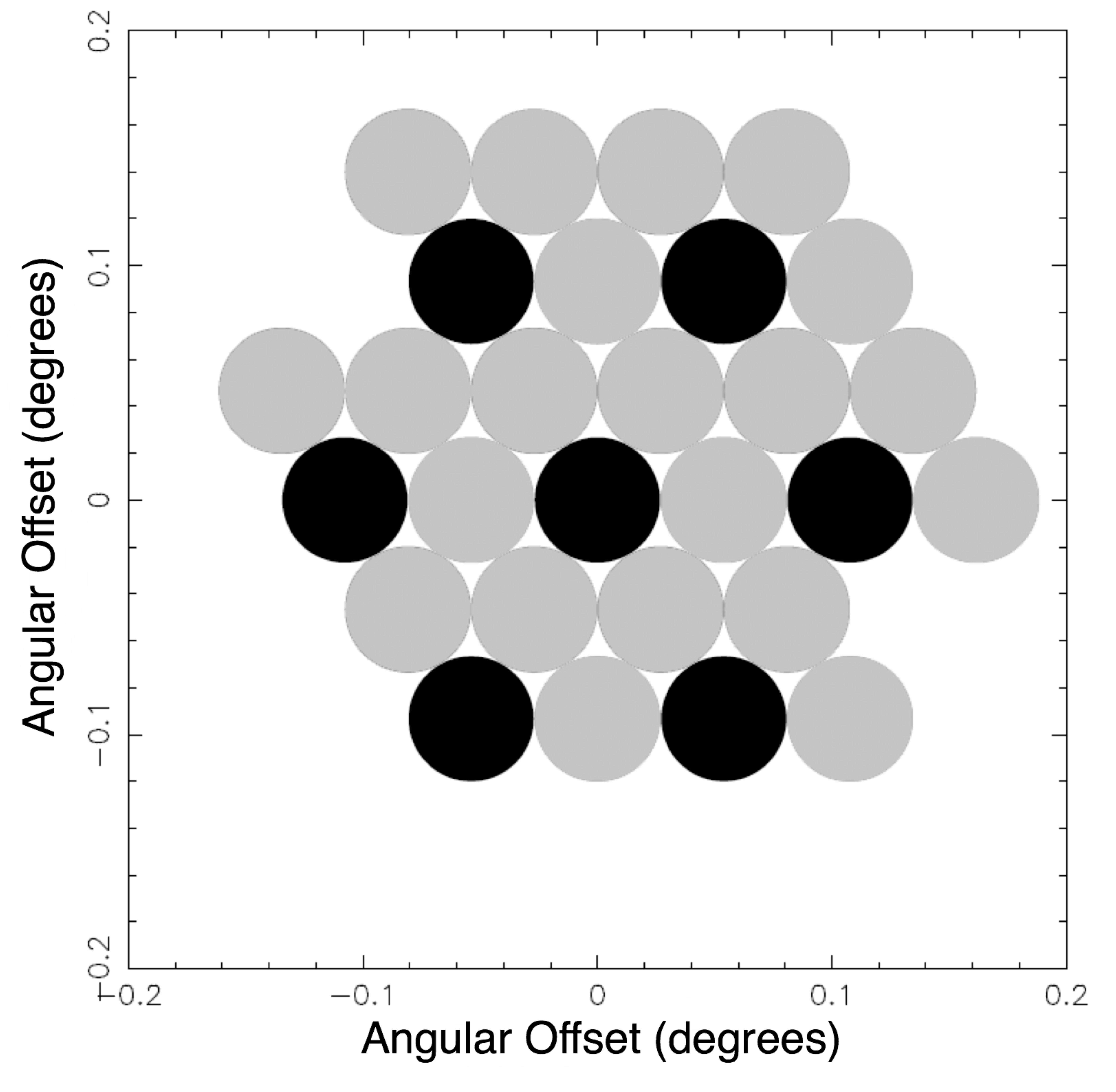}
	\end{center}
	\caption{Four pointings of the MMB receiver tessellated to give complete coverage over a whole patch of sky. One pointing is coloured black to show the pattern of the seven beams of the receiver, each of half-power width 0.053 degrees.}
	\label{beampattern}
\end{figure}

\section{Survey Equipment and Processing Systems}
\subsection{Collecting Data}
\begin{figure}
	\begin{center}	\includegraphics[width=8cm]{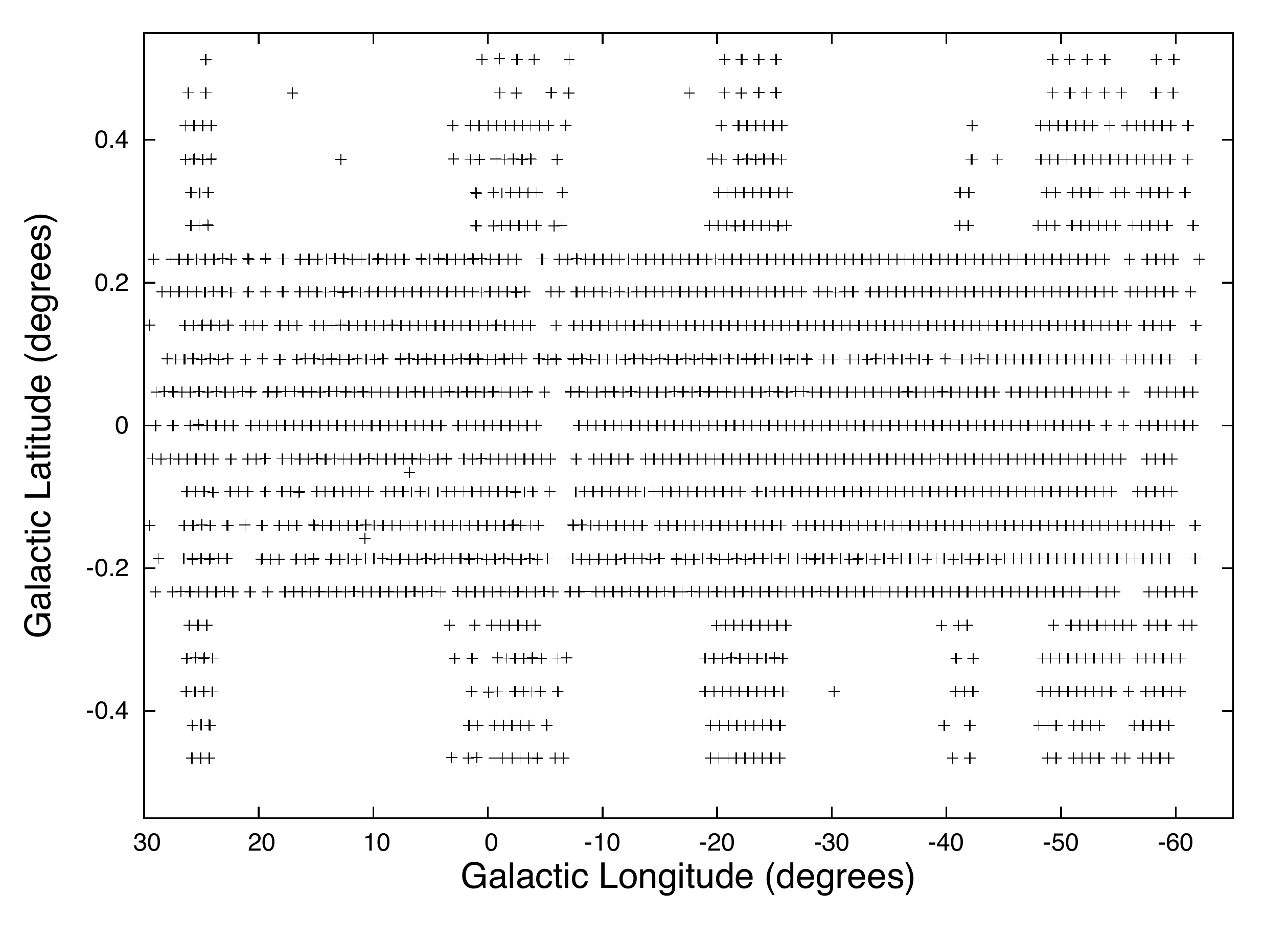}
	\end{center}
	\caption{All survey pointings for the Galactic-plane portion of the MMB survey plotted in Galactic coordinates. The dense coverage in the range $|b|\leq0.25\degree$ contrasts with that for $0.25\degree<|b|\leq0.5\degree$ where observations were not completed.}
	\label{skycoverage}
\end{figure}

\begin{figure}
	\begin{center}	\includegraphics[width=8cm]{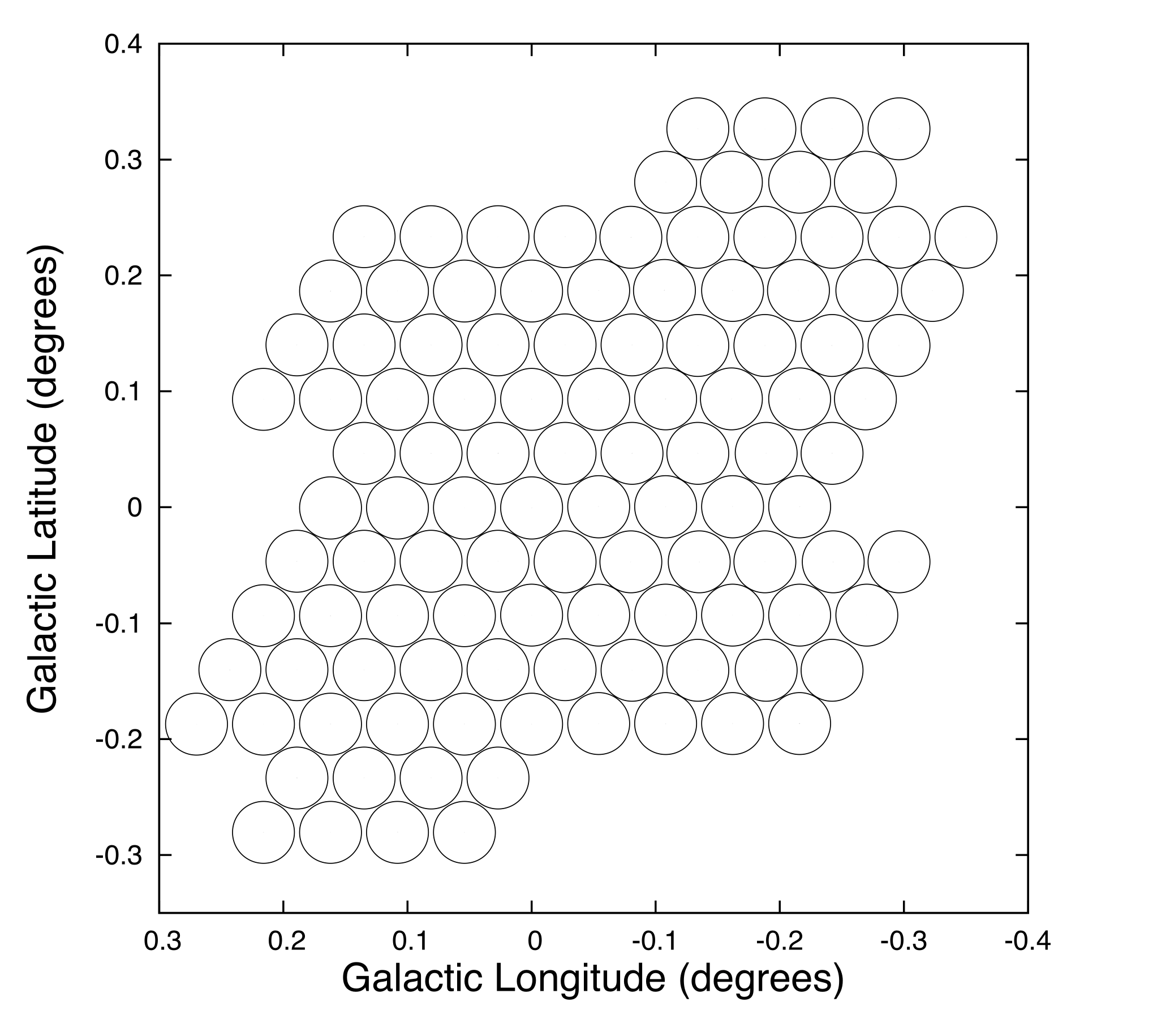}
	\end{center}
	\caption{The position of all observations from the Galactic-centre portion of the MMB survey, from tessellating the MMB receiver in 16 different positions.}
	\label{GCbeams}
\end{figure}

This survey utilised the seven-beam Methanol Multibeam (MMB) receiver at a central frequency of $6.5\,\mathrm{GHz}$ mounted upon the 64-metre Parkes radio telescope. The seven beams of the MMB can be tessellated in order to cover the entire survey region, as shown in Fig.~\ref{beampattern}, with each beam having a half-power width of 0.053 degrees. The original plan was to survey a thin strip of the Galactic plane from $-60\degree \leq l \leq 30\degree, |b| \leq 0.5\degree$. Subsequently, due to lack of success in finding previously unknown pulsars, the latitude range was reduced to $|b| \leq 0.25\degree$. Data were collected from February 2006 to September 2007. Table \ref{parameters} lists the survey parameters and Fig.~\ref{skycoverage} shows the area of sky covered by the survey observations, with each seven-beam pointing marked by a single cross. For $|b|\leq0.25\degree$ the survey is 85\% complete and for $0.25\degree<|b|\leq0.5\degree$ the completed fraction is 31\%.
In addition, as shown in Fig.~\ref{GCbeams}, a small area of $\pm0.3\degree$ around the Galactic centre was surveyed in July 2006 over 16 separate observations, with an integration time of 16,800 seconds per pointing, providing approximately four times the sensitivity of the observations of the Galactic plane region. Tables of the central beam positions in each survey are included in the supporting online material.

A large observing bandwidth of 576 MHz, covered by a series of 192 filterbank channels each of width 3 MHz, was used in order to counteract the effects of interstellar dispersion. This dispersion causes a time delay in seconds,
\begin{equation}\Delta\tau = 8.3 \times 10^3 \mathrm{DM}\nu^{-3}\Delta\nu\label{dispersion}\end{equation}
across an observing bandwidth of $\Delta\nu$ which is centred at frequency $\nu$ (both in MHz). During each pointing, the data were one-bit sampled every $125\,\mu \mathrm{s}$, written to a local computer disk at the Parkes site and, once the pointing was completed, were written to DLT-S4 tape. This amounted to $\sim$5~TB of data, which were subsequently processed offline at the Jodrell Bank Observatory, while a second copy of the survey was kept on tape at the the Australia Telescope National Facility for backup and for quick access to data in the event of discoveries.

\subsection{Survey Sensitivity}
The limiting flux density, in mJy, of a pulsar search \citep[see][]{dewey1985} can be calculated by
\begin{equation}
S_{\rm min}=\frac{(S/N)_{\rm min}\beta T_{\rm sys}}{G\sqrt{n_p t_{\rm obs} \Delta\nu}} \sqrt{\frac{\rm{W}_{\rm{eff}}}{\rm{P} - \rm{W}_{\rm{eff}}}}\label{senseq}.
\end{equation}
In this expression $(S/N)_{\rm min}$ is the minimum detectable signal-to-noise ratio,
$\beta$ is the degradation factor which takes into account effects such as signal digitisation, $T_{\rm sys}$ is the system temperature (defined as the sum of the sky, spillover and receiver temperatures), $G$ is the telescope gain, $n_{\rm p}$ is the number of polarisations summed ($n_p = 2$), $t_{\rm obs}$ is the observation time and $\Delta\nu$ is the bandwidth in MHz. P, the pulse period, and W$_{\rm eff}$, the effective pulse width as defined in equation~(\ref{weffeq}), are properties of the pulsar to be detected, rather than survey parameters.

\begin{figure}
	\begin{center}	\includegraphics[width=8cm]{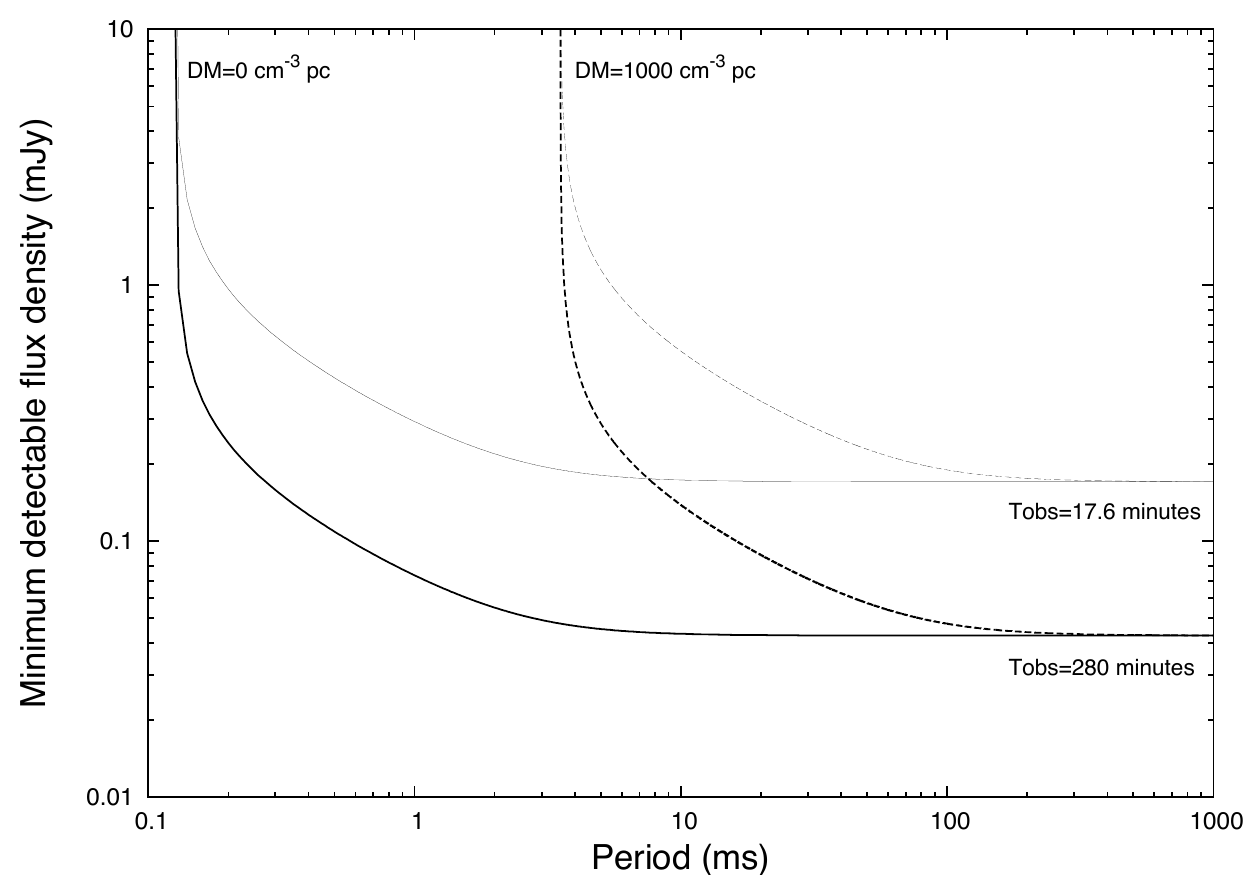}
	\end{center}
	\caption{Minimum detectable flux density (in mJy) of a source as a function of the period (in ms) for a detection at the 8$\sigma$ level in  the MMB survey when placed at a DM of 0 and 1000~$\mathrm{cm}^{-3}\,\mathrm{pc}$, calculated using equation~(\ref{senseq}). The sensitivities were calculated for integration times of 17.6 minutes and 280 minutes, for the Galactic-plane and Galactic-centre surveys respectively.}
	\label{sens_plot}
\end{figure}

Fig.~\ref{sens_plot} shows the computed sensitivity curves for dispersion measures of 0 and $1000\,\mathrm{cm}^{-3}\,\mathrm{pc}$ respectively, using a mean sky temperature of 0.5~K, assuming $S/N_{\rm min}=8$. The intrinsic width of the pulse is assumed to be 5\% and we use the scattering model of \citet{bcc+04}, shown in equation~(\ref{bhatscatter}). The curves indicate that at periods greater than about 10 ms, the sensitivity is approximately 0.15 mJy. It is also evident that the large change in DM between the two curves results in a very small change in limiting sensitivity at periods above $\sim$100~ms, which is an effect of the high observing frequency used in this survey, discussed further in Section \ref{sec:high_f}.

% here I edit from sky temp in Jy to K - multiplied through by the Gain (0.6 K Jy^-1)
For the Galactic-centre region, the sensitivity is affected by the background temperature of the Galactic centre itself. Using the maps of \citet{seiradakis1989}, we estimate this contribution to be $\sim$4~K in the outer regions of the survey, $\sim$12~K in the inner regions and $\sim$90~K at the GC. The detection threshold of $S/N_{\rm min}=9$ is then $\sim$28~$\mu$Jy (outer regions), $\sim$32~$\mu$Jy (inner regions) and $\sim$81~$\mu$Jy at the GC.

\subsection{Processing the data - the first pass}
The processing pipeline used the SIGPROC-4.2 software package\footnote{http://sigproc.sourceforge.net/} to perform dedispersion of the data up to a DM of $2000\,\mathrm{cm}^{-3}\,\mathrm{pc}$ --- chosen to ensure the survey was capable of observing pulsars at the Galactic centre --- using a DM step size of $7.18\,\mathrm{cm}^{-3}\,\mathrm{pc}$, derived from equation~(\ref{dispersion}), ensuring that the residual dispersive time delay between the highest and lowest frequency channels is no greater than one sample. These dedispersed time series were fast Fourier-transformed (FFT), and the spectrum was searched for significant peaks, as were spectra with 2, 4, 8 and 16 harmonics summed in order to ensure power was obtained from narrow pulses \citep{mld+95}. The Pulsar Hunter package\footnote{http://pulsarhunter.sourceforge.net} was then used to search in period and DM space around the values output from the FFT search, in order to maximise the S/N ratio of each candidate.

The pipeline was run on `DCore', a beowulf cluster with 72 processors, and candidates were inspected with JReaper \citep{kel+09}, a graphical tool for selecting candidates based upon user-definable parameters. This first pass yielded only one new discovery, J1410$-$6132 \citep{obrien08}, discussed in Section 4.1.

One of the possible reasons why the first pass through the data found only one pulsar may have been the lack of dispersion delay, making it difficult to distinguish astrophysical signals from RFI at this high frequency (as discussed in Section \ref{sec:high_f}). This resulted in an overwhelming number of spurious candidates, making the task of finding good candidates which should be reobserved very difficult. Therefore, it was decided that a second pass through the data would be useful, after introducing some techniques to remove, or reduce the effects of, these RFI signals.

\subsection{Reprocessing the data}\label{sec:reprocessing}
In reprocessing the data, much of the pipeline was very similar to that used originally. However, the processing was carried out on `Hydra', a 108-node cluster, with each node consisting of dual quad-core processors with 4GB of memory. In order to take advantage of this processing power, a threaded tree algorithm written by one of us (Bailes) was used, which performed dedispersion many times faster than was possible with `DCore' and the original dedispersion code.

When preparing the data for processing, we decided not to use the time-domain clipping algorithm that was previously employed by this and other surveys, since the discovery that this algorithm introduced periodic signals to the data. These signals were apparent even at high DM, polluting the search output. By removing this step, the effects of RFI at low DM were increased, however the sensitivity to these artificial signals was removed. Steps were taken to reduce our sensitivity to RFI in other ways, outlined in the following sections.

Instead of making changes to the processing pipeline, alterations were made in the post-processing stage, as outlined in Section \ref{sec:filters}. For example, the data were not searched in acceleration space to look for pulsars in short-period binary systems. The additional computation required by such a search is large, while this technique would also produce enormous numbers of candidates, something that we were attempting to avoid with this reprocessing.

\section{Searching for pulsars at high frequencies}

\subsection{Dispersion as a discriminatory tool}\label{sec:high_f}
\begin{figure}
	\begin{center}	\includegraphics[width=8cm]{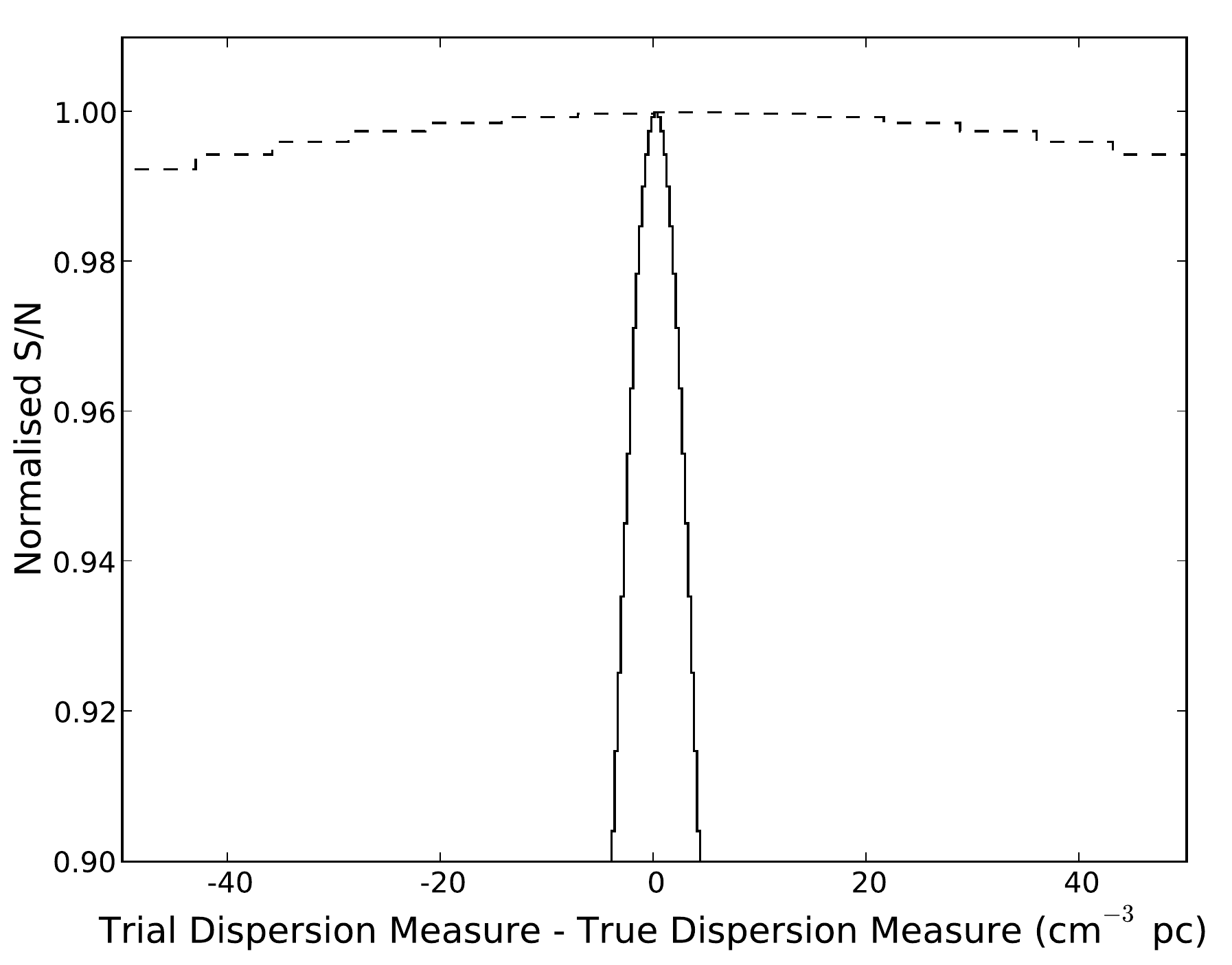}
	\end{center}
	\caption{The normalised signal-to-noise ratio of a periodic signal, with a duty cycle of 5\% and period 500~ms, as the difference between the true DM value and the trial DM value varies. This is shown for both the MMB survey (dashed line) and the PMPS (solid line). The DM step sizes shown are those used for processing data in the respective surveys.}
	\label{snr_vs_dm_f}
\end{figure}

Interstellar dispersion, described by equation~(\ref{dispersion}), is often a good way to distinguish between extra-terrestrial sources and terrestrial sources (that is, RFI), since one would ordinarily expect that any periodic sources of RFI would not follow the dispersion law, and that they would peak in strength at a DM of 0~$\mathrm{cm}^{-3}\,\mathrm{pc}$.
At high frequencies, however, where the effects of dispersion are small, a signal with zero dispersion will reduce in S/N very slowly as trial DM increases - the simulated pulse in Fig.~\ref{snr_vs_dm_f} falls in S/N by less then one percent when the difference between the true DM and the trial DM is 50~$\mathrm{cm}^{-3}\,\mathrm{pc}$. By comparison, at an observing frequency of 1.4~GHz (in, for example, the PMPS), the same simulated pulse falls by ten percent in S/N when the true and trial DMs differ by only 4~$\mathrm{cm}^{-3}\,\mathrm{pc}$. Due to their short periods, this effect should not reduce sensitivity to millisecond pulsars (MSPs), for whom the dependance of S/N with DM is very strong.

The effect of the small amount of dispersion delay at high frequencies is that the variation of S/N with trial DM value is low compared to observations at lower frequencies. It is normal to use a plot of the variation of S/N with DM to help identify good candidates, but for this survey these plots were of little value, except for the highest DM sources (see Section \ref{sec:plane}). In lower DM cases, one has to judge instead based purely upon whether the periodic signal is persistent in time and frequency, and upon the shape of the pulse profile, with little or no indication of whether a source is actually terrestrial in origin. This led to there being an overwhelming number of candidates, especially at low DM. A typical single beam from the survey would generate $\sim$170 candidates, leading to a total of over 3.5 million candidates in the survey.

\subsection{Reducing the number of candidates}\label{sec:filters}
This large number of candidates is impossible to filter by eye and so some automated filters were designed to reduce the number to be viewed. As the MMB receiver has multiple beams, it allows us to compare each of the beams from the same observation, and assume that if a periodic signal is detected in many beams, then it is likely to be ground-based RFI. Therefore, any periodicities which appeared in two or more beams  from the same pointing and had a period which matched within a tolerance of 10~ns were rejected.

As shown in equation~(\ref{senseq}), the S/N of a periodic signal decreases as the effective pulse width increases for a given set of observing parameters. The effective width is given by
\begin{equation}
W_{\mathrm{eff}}^2=W_{\mathrm{int}}^2+W_{\mathrm{samp}}^2+\Delta t^2+\tau_\mathrm{s}^2\label{weffeq}
\end{equation}
where $W_{\mathrm{samp}}$ is equal to the sampling rate and $\tau_\mathrm{s}$ is a distortion in the pulse width due to scattering in the ISM (given by equation~(\ref{bhatscatter})). Equation~(\ref{dispersion}) can be used to give $\Delta t$, where $\Delta\nu$ is now the bandwidth of a single filterbank channel. This broadening of the pulse arises due to the dedispersion process and the finite width of the filterbank channels in the observing system.

\begin{figure}
	\begin{center}
	\includegraphics[width=8cm]{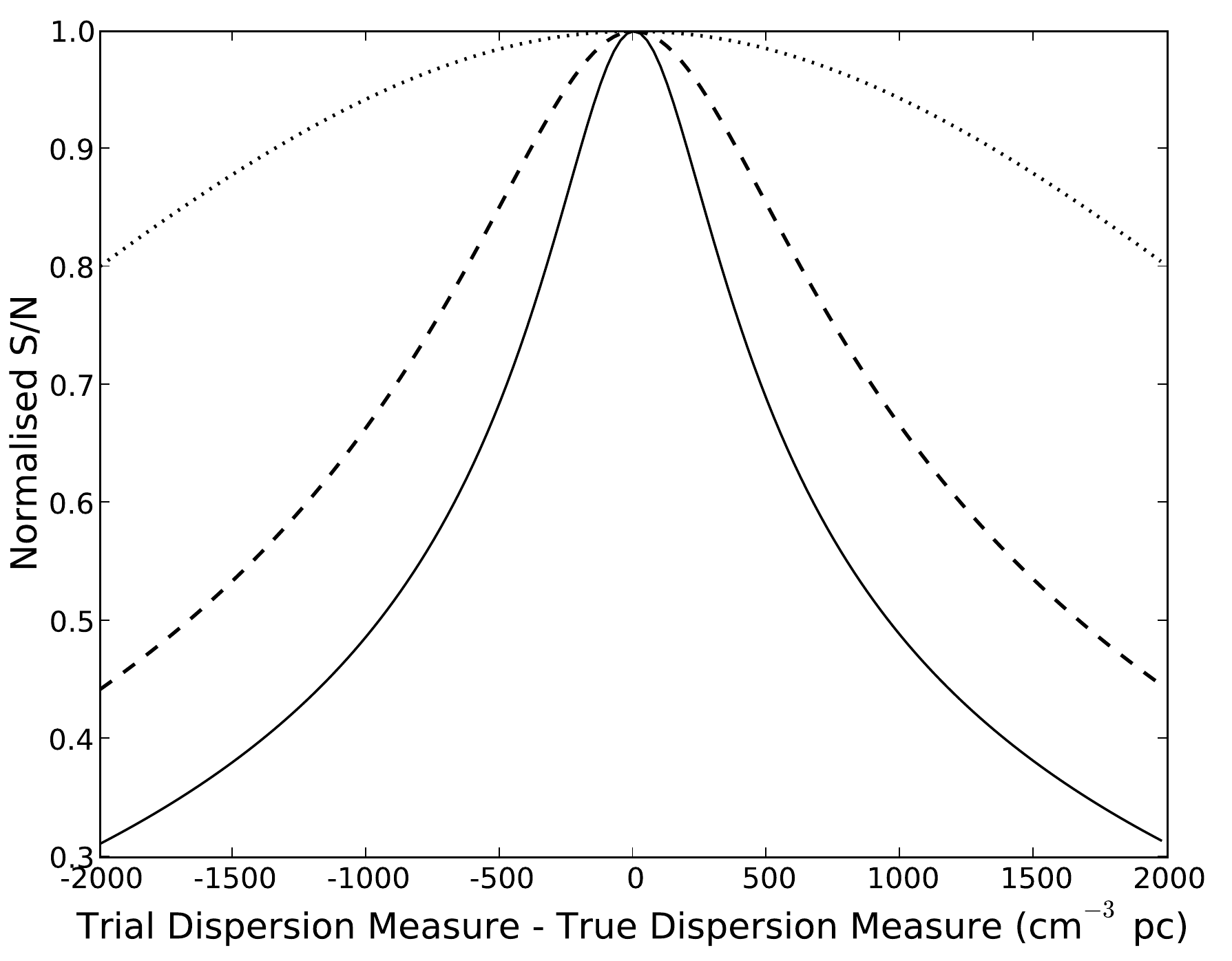}
	\end{center}
	\caption{The normalised signal-to-noise ratio of a signal with period 100~ms, for trial DM values every $25\,\mathrm{~cm}^{-3}\,\mathrm{pc}$, at the MMB observing frequency of 6.5~GHz. The dotted, dashed and solid lines show S/N for signals with W/P 50\%, 10\% and 5\% respectively.}
	\label{snr_vs_dm_w}
\end{figure}

Fig.~\ref{snr_vs_dm_w} shows S/N against trial DM error for signals with duty cycles of 50\%, 10\% and 5\%; it can be seen that S/N drops off more quickly for narrower pulses. Thus, a filter was applied to the candidates to reject any with a width greater than 10\% of the period, which eliminated sinusoidal signals often indicative of RFI, which are often of high S/N for a very large range of trial DMs. In total, these two filters removed $\sim$15\% of the candidates from the processing pipeline.

During the inspection of the candidates, a further cut was made, removing all candidates with S/N $<$ 9. This removed a further 50\% of the candidates, with many of those remaining clustered around common RFI frequencies. It is common, during the inspection process, to visualise the candidates in Period-DM space (for an example using the JReaper software, see Fig.~2 in \citet{kel+09}), so that any `clustering' of the candidates at RFI frequencies is highly apparent, making it easier for such candidates to be ignored. This reduced the number of candidates to $\sim10000$, and given that it takes only a few seconds to classify a candidate, this allowed visual inspection of this subset of the survey output.

\section{Results \& Discussion}
\subsection{New Discoveries}\label{sec:newpsrs}
\subsubsection{Galactic Plane Region}\label{sec:plane}
The first processing of the data, by \citet{obrien08}, uncovered one previously unknown pulsar, the young and highly energetic PSR J1410$-$6132. \citet{obrien08} noted that the pulsar was within the error box of the EGRET source 3EG J1410$-$6147 and it is within that of the Fermi source J1410.3$-$6128c \citep{abdoetal}. The pulsar is regularly observed at Parkes as part of the timing support for the Fermi mission \citep{weltevrede2010}, but as yet pulsations in gamma-rays have not been detected \citep{fermipsrcat}.

\begin{figure}
	\begin{center}	\includegraphics[width=8cm]{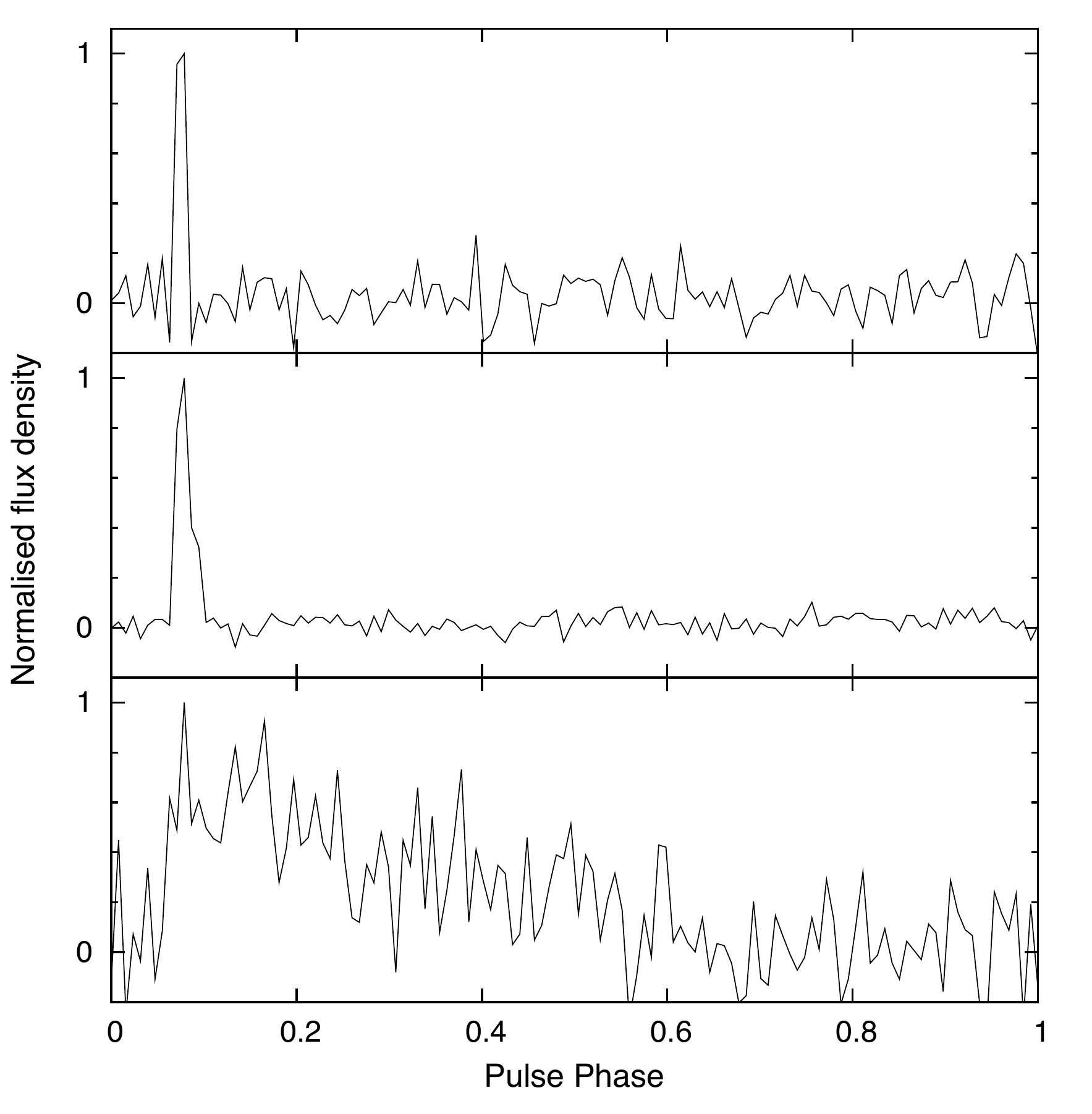}
	\end{center}
	\caption{Pulse profiles of PSR J1834$-$0812 at observing frequencies of 6.5~GHz (top panel), 4.85~GHz (middle panel) and 1.4~GHz (bottom panel), with the peak of each profile positioned at phase 0.1. The increase in pulse width as observing frequency is reduced, due to scattering, is clearly visible, particularly at 1.4~GHz.}
	\label{1834prof}
\end{figure}

\begin{figure}
	\begin{center}	\includegraphics[width=8cm]{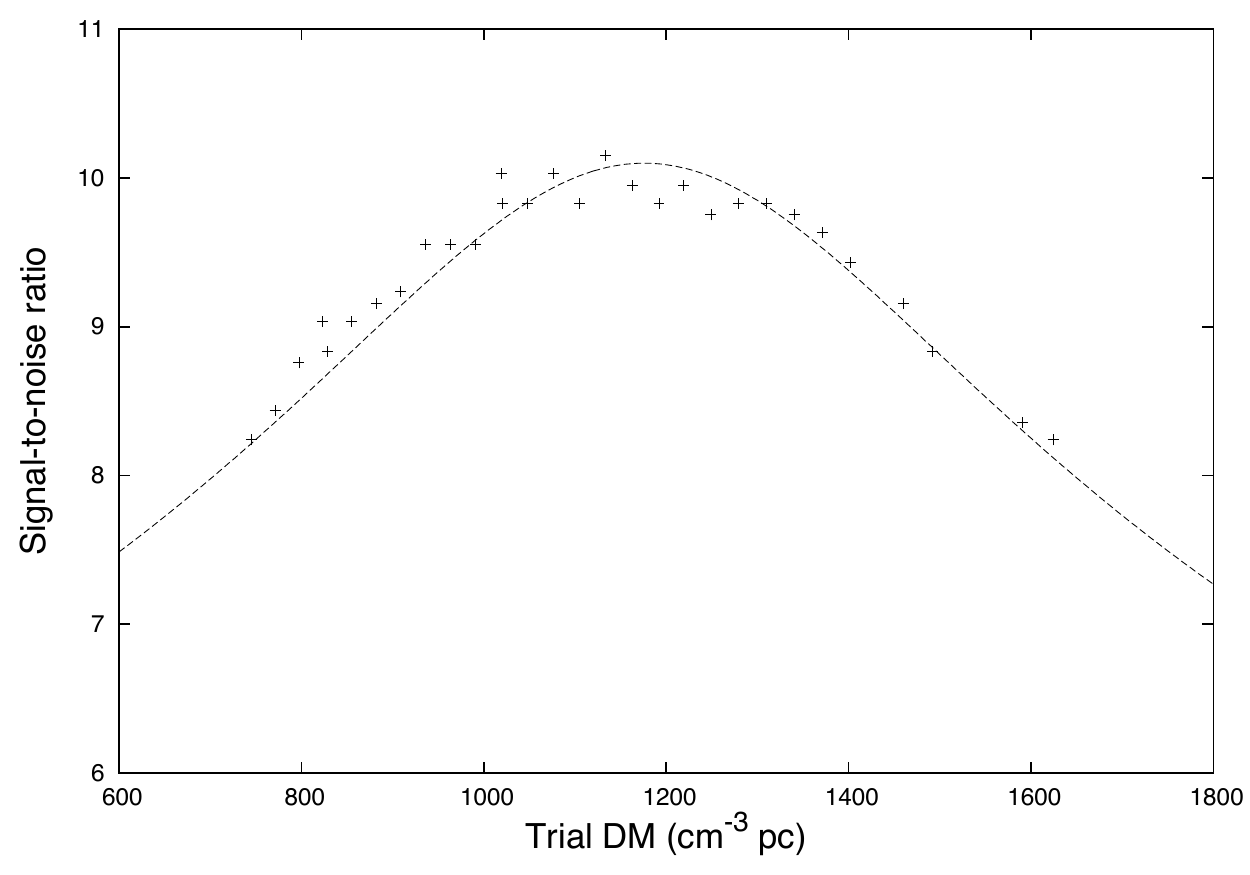}
	\end{center}
	\caption{S/N of PSR J1834$-$0812 for different trial DM values used in the data processing pipeline. Note that the processing pipeline discovery slightly over-estimated the DM, hence the peak is positioned here at $1175\,\mathrm{cm}^{-3}\,\mathrm{pc}$}
	\label{1834snrdm}
\end{figure}

One further pulsar, PSR J1834$-$0812, with a period of 491~ms and a DM of $1023\,\mathrm{cm}^{-3}\,\mathrm{pc}$ was discovered during the reprocessing of the Galactic-plane data (detailed in Section \ref{sec:reprocessing}). As seen in the top panel of Fig.~\ref{1834prof}, the pulse profile at 6.5~GHz is narrow with a duty cycle of only 2\%, which ensured that the S/N was significantly lower away from the true value of DM, as shown in Fig.~\ref{1834snrdm}.

\begin{table*}
	\begin{center}
	\caption{Measured and derived parameters for each of the pulsars discovered in the survey. Numbers in parentheses signify the error in the last digit. Values for J1746$-$2850 are taken from the timing solution of \citet{deneva2009}, while those for J1834$-$0812 are based on an interim timing solution spanning $\sim$150 days.}
		\begin{tabular}{lcc}
		\toprule
		Pulsar Name & J1746$-$2850 & J1834$-$0812\\
		\midrule
		%\multicolumn{2}{c}{Receiver Parameters} \\
		Right Ascension (J2000) & 17:46:06.6(2) & 18:34:29.8(9)\\
		Declination (J2000) & $-$28:50:42(5) & $-$08:12:00(100)\\
%		Galactic Longitude (degrees) & 0.134 & 23.7\\
%		Galactic Latitude (degrees) & $-$0.0441 & 0.00172 \\
		Period (s) & 1.0771014910(4) & 0.4911415563(2)\\
		Period Derivative & 1.34311(2) $\times 10^{-12}$ & 9.96(7) $\times 10^{-15}$\\
		Dispersion Measure ($\mathrm{cm}^{-3}\,\mathrm{pc}$) & 962.7(7) & 1020(50) \\
		\\
		Flux Density at 6.5~GHz (mJy) & 0.14(3) & 0.10(2) \\
		Spectral Index & $-0.3$ & $-0.7$\\
		DM Distance (kpc) & 11.6$_{-2.1}^{+1.2}$ & 9.6(1.2)\\
		%\multicolumn{2}{c}{Survey Parameters} \\
		\bottomrule
		\end{tabular}
		\label{psrpars}
	\end{center}
\end{table*}

The pulsar has been observed at a frequency of $4.85\,\mathrm{GHz}$ using the Effelsberg 100-m radio telescope, and at 1.4~GHz with the Lovell telescope. The profile is extremely scatter broadened at this frequency, with a scattering time of $\sim$150~ms. The pulse profiles at 4.85 and 1.4~GHz are shown in the bottom two panels of Fig.~\ref{1834prof}.

\subsubsection{Galactic Centre Region}
\begin{figure}
	\begin{center}	\includegraphics[width=8cm]{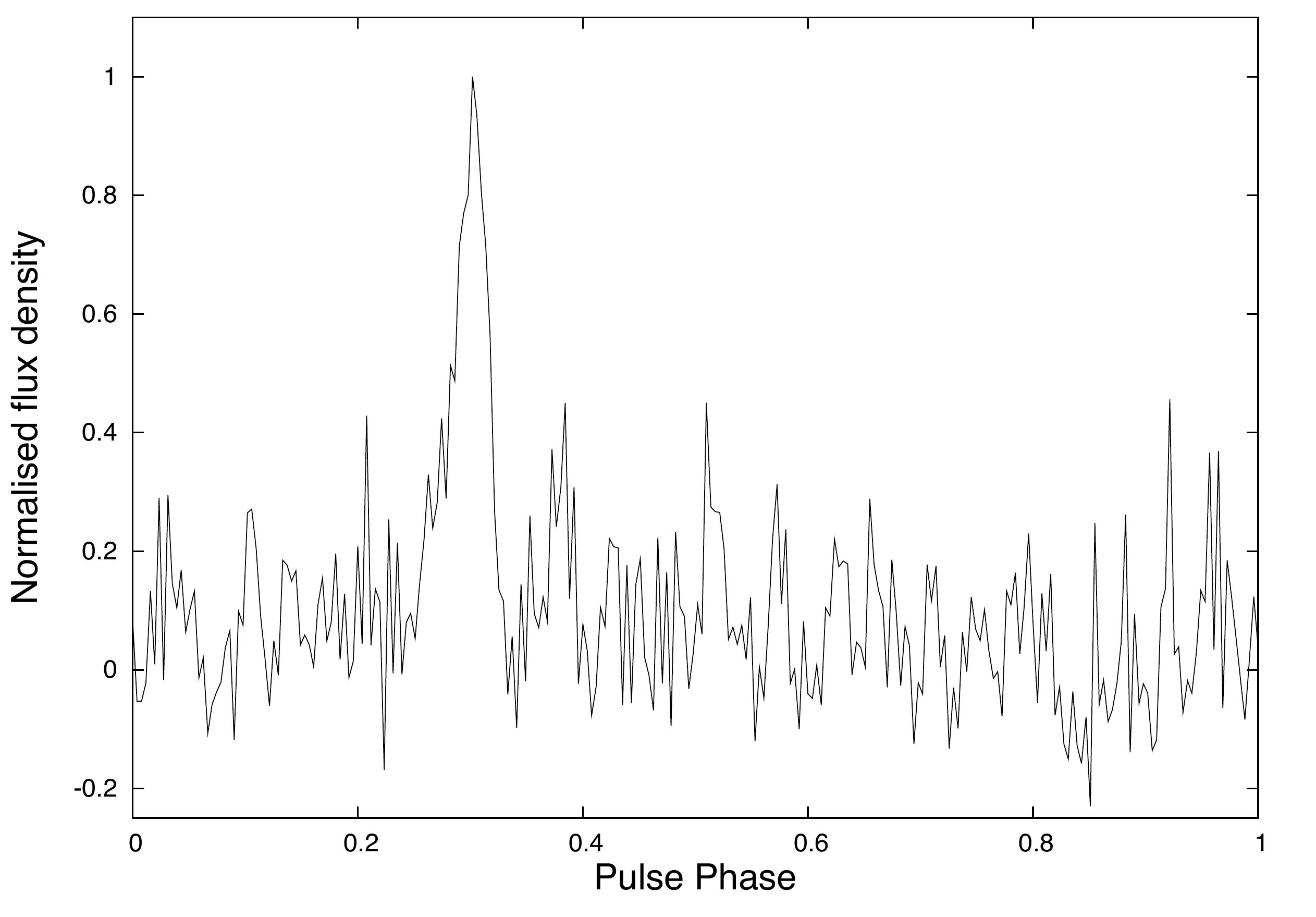}
	\end{center}
	\caption{Pulse profile of PSR J1746$-$2850, discovered in the Galactic-centre portion of the MMB pulsar survey.}
	\label{1746prof}
\end{figure}

Processing of the data taken around the Galactic centre was completed in August 2006 and resulted in an excellent candidate. The candidate was subsequently confirmed to be a pulsar, PSR J1746$-$2850, with a period of $1077\,\mathrm{ms}$ and a DM of $963\,\mathrm{cm}^{-3}\,\mathrm{pc}$ in an observation made on October 29, 2006. Subsequently, this pulsar was also seen by \citet{deneva2009}, who have published a full timing solution. A folded pulse profile of this pulsar from a confirmation observation taken with the MMB receiver is shown in Fig.~\ref{1746prof}.

Parameters for PSRs J1746$-$2850 and J1834$-$0812 are shown in Table~\ref{psrpars}, including DM distances estimated using the Galactic electron distribution model of \citet{ne2001}. As the parameters for PSR J1834$-$0812 are based on timing data that does not yet span one year, increased precision will be obtained with further observations and as period derivative and position become non-covariant.

\subsection{Known Pulsars}
Within the survey region, there are 113 previously known pulsars. Of these 113, 16 were detected. 

\subsubsection{Non-detections}
A total of 62 pulsars fall below our detection threshold (taking the offset from the beam centre into account) for any sensible value of the spectral index (i.e. where $\alpha <-0.5$). For the remaining 16 pulsars, we can compute an upper bound on the spectral index by using the flux density at 1.4~GHz, as given in the pulsar catalogue\footnote{http://www.atnf.csiro.au/research/pulsar/psrcat} \citep{mhth05}. An upper limit to the 6.5~GHz flux density was calculated assuming a sensitivity of 0.2~mJy at the beam centre and the half power beam width of 0.053 degrees. These calculated values are shown in Table~\ref{nondetections}.

Of the limits set on the spectral index, the steepest is set at $-2.0$, for PSR J1809$-$1943. This would not be an unusually steep value for the spectral index, well within the estimated distributions of both \citet{maron2000} and \citet{lorimer1995}. Therefore, we can be confident that the flux densities are too low to have made them visible in the survey data.

\subsubsection{Detections}
Only one of the pulsars, PSR J1746$-$2856, was undetected in the processing but later found when folded. This is likely the result of strong RFI during that observation.

For those pulsars that were detected, Table~\ref{fluxpred} shows the S/N from the search pipeline (S/N Search), and that obtained when the data were folded at the period of the pulsar (S/N Fold). The other columns show the flux density at 1.4~GHz \citep{lorimer1995}, 4.85~GHz (obtained during the work of \citet{maron2000}) and as estimated from our survey detections, using equation~(\ref{senseq}).

We find that there is a large range of values for the spectral index, between $0.03$ and $-2.30$ with an average of $-0.94$. This value is shallower than that found by \citet{maron2000} who investigated pulsar spectra between 0.4 and 4.9~GHz. However, by observing at a high frequency for a fixed duration, we are biased to observing pulsars which have a flat spectral index, or that are sufficiently bright to be observable at this high frequency. Since the majority of pulsars have flux densities of a few mJy or less at 1.4~GHz, it is not unexpected that our sample should contain many pulsars with a shallow spectral index.

\subsubsection{PSR J1550$-$5418}
PSR J1550$-$5418 is a radio-emitting magnetar \citep{cam07}. The pointing containing this source was taken on April 23rd 2007, 46 days prior to the observation by \citet{cam08} in which PSR J1550$-$5418 was discovered. Unfortunately, the data were not passed through the pipeline until after the announcement of the discovery, in the radio at 1.4~GHz, by \citet{cam07}. The flux density calculated from our observation does, however, agree with the observations of \citet{cam08}, given both the lack of a flux density  calibrator for this single observation and the time variation of the flux density for that source. Since our observation fell between the initial observations, in the X-ray, of \citet{gg2007} and the radio detection of \citet{cam08}, we are only able to conclude that radio emission commenced at least 46 days before the published discovery.

\begin{table}
	\caption{Basic parameters and flux densities at 1.4~GHz for all undetected pulsars in the survey region with a limit on the spectral index steeper than $-0.5$. The last column, $\alpha_{\rm lim}$, gives an upper limit for the spectral index. It is calculated assuming a maximum flux density at 6.5~GHz of 0.2~mJy for a source at the centre of the survey beam.}
		\begin{tabular}{ccccc}
		\toprule
		Pulsar Name & Period & DM & $S_\mathrm{1400}$ & $\alpha_\mathrm{lim}$ \\
		 (J2000) & (seconds) & ($\mathrm{cm}^{-3}\,\mathrm{pc}$) & (mJy) & \\
		\midrule
J1512$-$5759 & 0.128694 & 628.70 & 6.0(6) & $-$1.7(1)\\
J1614$-$5048 & 0.231694 & 582.80 & 2.4(3) & $-$1.5(1) \\
J1648$-$4458 & 0.629632 & 925.00 & 0.55(7) & $-$0.6(1) \\
J1705$-$4108 & 0.861067 & 1077.00 & 1.3(2) & $-$1.0(1) \\
J1717$-$3737 & 0.682419 & 525.80 & 0.69(8) & $-$0.76(13) \\ \\
J1722$-$3632 & 0.399183 & 416.20 & 1.60(17) & $-$1.1(1) \\
J1730$-$3350 & 0.139460 & 259.00 & 3.2(3) & $-$1.8(1) \\
J1739$-$3131 & 0.529441 & 600.10 & 4.9(5) & $-$1.4(1) \\
J1740$-$3052 & 0.570310 & 740.90 & 0.7(2) & $-$0.8(3) \\
J1741$-$3016 & 1.893749 & 382.00 & 2.3(5) & $-$0.6(2) \\ \\
J1756$-$2435 & 0.670480 & 367.10 & 2.0(2) & $-$1.5(1) \\
J1809$-$1943 & 5.540249 & 178.00 & 4.5(5) & $-$2.0(1) \\
J1818$-$1519 & 0.939690 & 845.00 & 2.1(4) & $-$1.4(2) \\
J1818$-$1541 & 0.551134 & 690.00 & 1.0(2) & $-$0.9(2) \\
J1819$-$1510 & 0.226539 & 421.70 & 0.6(1) & $-$0.6(2) \\ \\
J1822$-$1400 & 0.214771 & 651.10 & 0.80(9) & $-$0.6(1) \\
		\bottomrule
		\end{tabular}
		\label{nondetections}
\end{table}

\begin{table*}
	\caption{S/N values and flux density at 1.4, 4.85 and 6.5~GHz, and the computed spectral index (from 1.4 to 6.5~GHz) for the pulsars detected in the survey data. The final column shows the measured scattering timescale at 1.4~GHz and its error.}
		\begin{tabular}{cccccccccc}
		\toprule
		Pulsar Name & Period & DM & S/N & S/N & S$_{1.4}$ & S$_{4.85}$ & S$_{6.5}$ & $\alpha$ & $\tau_\mathrm{s}$ \\
		 (J2000) & (seconds) & ($\mathrm{cm}^{-3}\,\mathrm{pc}$) & Fold & Spectral & (mJy) & (mJy) & (mJy) & & (ms) \\
		\midrule
J1316$-$6232 & 0.342825 & 983.3 & 41.5 & 21.5 & 0.74(8) & --- & 0.7(1) & $-0.0(1)$ & 29(8)\\
J1327$-$6222 & 0.529913 & 318.8 & 39.4 & 39.5 & 16.0(1.6) & --- & 0.9(2) & $-1.9(1)$ & ---\\
J1341$-$6220 & 0.193340 & 717.3 & 22.2 & 13.5 & 1.9(2) & --- & 0.8(2) & $-0.5(1)$ & 8.3(3)\\
J1406$-$6121 & 0.213075 & 542.3 & 13.8 & 11.1 & 0.36(5) & --- & 0.20(5) & $-0.4(2)$ & 9(1)\\
J1410$-$6132 & 0.050052 & 960.0 & 12.4 & 7.7 & 6(1) & --- & 0.6(1) & $-1.5(2)$ & 33(15)\\ \\
J1531$-$5610 & 0.084202 & 110.9 & 15.6 & 15.4 & 0.60(7) & --- & 0.4(1) & $-0.2(2)$ & --- \\
J1550$-$5418 & 2.069833 & 830.0 & 231 & 199 & 3.3(3) & --- & 9(2) & $+0.7(1)$ & --- \\
J1644$-$4559 & 0.455060 & 478.8 & 703 & 475 & 310 & --- & 12(2) & $-2.1(1)$ & 2.45(2)\\
J1702$-$4128 & 0.182136 & 367.1 & 25.6 & 24.3 & 1.10(12) & --- & 0.6(1) & $-0.4(1)$ & 7.5(4)\\
J1707$-$4053 & 0.581017 & 360.0 & 10.7 & 10.3 & 7.2(7) & --- & 0.19(5) & $-2.3(1)$ & 26.1(9)\\ \\
J1740$-$3015 & 0.606784 & 152.2 & 18.5 & 11.2 & 6.4(7) & 1.08(0) & 1.7(3) & $-0.9(1)$ & ---\\
J1746$-$2856 & 0.945224 & 1168 & 11.01 & N/A & --- & --- & 0.10(2) & --- & --- \\
J1757$-$2421 & 0.234101 & 179.5 & 15.1 & 16.3 & 3.9(4) & --- & 1.0(2) & $-0.9(2)$ & --- \\
J1801$-$2304 & 0.415796 & 1074 & 24.1 & 16.8 & 2.2(2) & 0.27(3) & 0.5(1) & $-1.0(1)$ & 99(9) \\
J1803$-$2137 & 0.133617 & 234.0 & 85.4 & 81.7 & 7.6(8) & 5.5(1.9) & 3.1(6) & $-0.6(1)$ & 5.5(7) \\ \\
J1809$-$1917 & 0.082747 & 197.1 & 36.7 & 35.7 & 2.5(5) & --- & 1.6(3) & $-0.3(2)$ & --- \\
J1832$-$0827 & 0.647293 & 300.9 & 14.6 & 13.0 & 2.1(2) & 0.14(0) & 0.20(5) & $-1.7(1)$ & --- \\
J1834$-$0812 & 0.491101 & 1023 & 18.0 & 12.0 & 0.3(1) & --- & 0.11(2) & $-0.7(3)$ & 150(100) \\
		\bottomrule
		\end{tabular}
		\label{fluxpred}
\end{table*}

\subsection{Spectral index distribution}\label{sec:SID}
We can check for the completeness of our survey in the following way. According to \citet{maron2000}, the spectral index distribution is Gaussian, with a mean of $-1.8$ and a standard deviation of $0.2$, whereas in \citet{lorimer1995} the mean is $-1.6$ and the standard deviation $0.3$. We take each of the 113 previously-known pulsars in our survey region (gathered using the pulsar catalogue, \citet{mhth05}) with a known flux densities at 1.4~GHz (leaving 103 in the sample), and compute the probability of their detection given the Maron and Lorimer distributions.

To do this we choose a pulsar at random from the 103 pulsars which remain, and assign it a spectral index chosen at random from a gaussian distribution with mean and standard deviation given by the Maron and Lorimer distributions. Using this spectral index, we then scale the flux density at 1.4~GHz (given by the pulsar catalogue, \citet{mhth05}) to 6.5~GHz and infer whether this pulsar would be above the detection threshold of 0.2~mJy. Repeating many times and normalising the results to 115 pulsars (since two have been discovered), we obtain an expected number of detected pulsars for each distribution.

The expectation is that we should have found 12.6 pulsars for the Maron et al. distribution and 16.1 for the Lorimer et al. distribution. The fact that we detected 18 indicates that 
\begin{inparaenum}[\itshape a\upshape)]
\item our survey is reasonably complete; and
\item there is evidence for there being a small excess of pulsars with flatter spectral indices than predicted by either \citet{lorimer1995} or \citet{maron2000}, but no significant population with extremely shallow spectral indices.
\end{inparaenum}

\subsection{Scattering}
By taking the 115 known pulsars (including our discoveries) in the survey region, it is possible to analyse the impact of scattering by the ISM. This is especially important at low Galactic latitudes, where scattering is most severe and which was not well covered by the \citet{bcc+04} survey. Scattering by the ISM causes a pulse profile to be convolved with an exponential tail of decay timescale $\tau_\mathrm{s}$, which was found to follow the empirical relation (for $\tau_\mathrm{s}$ in milliseconds),
\begin{equation}
\log\tau_\mathrm{s} = -6.46+0.154\log\mathrm{DM}+1.07(\log\mathrm{DM})^2-3.86\log{\nu},\label{bhatscatter}
\end{equation}
at a frequency of $\nu$~GHz, for a source with dispersion measure DM$\,\mathrm{cm}^{-3}\,\mathrm{pc}$ \citep{bcc+04}.

\begin{figure}
	\begin{center}	\includegraphics[width=8cm]{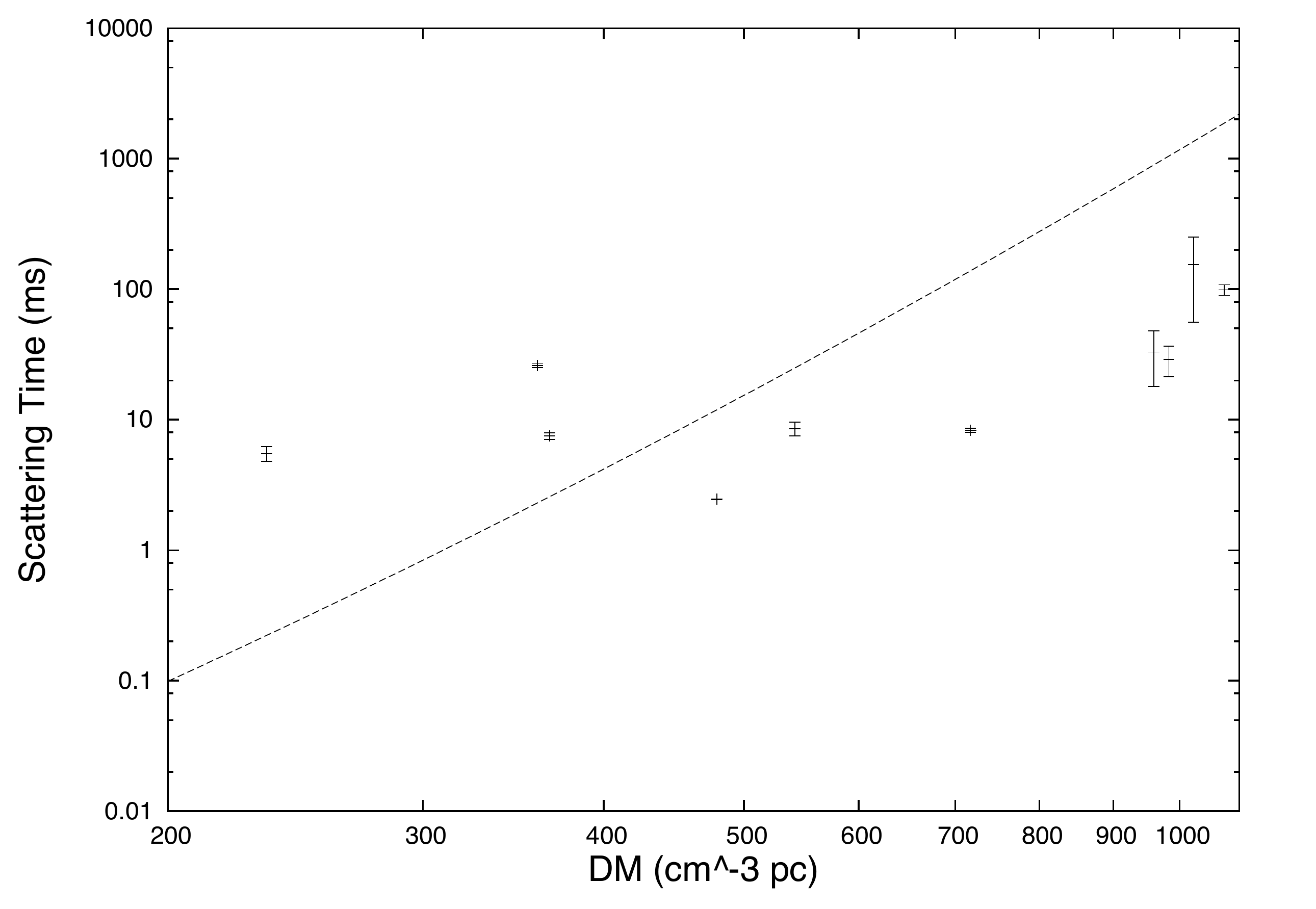}
	\end{center}
	\caption{Characteristic width of an exponential scattering tail convolved with a pulse profile at 6.5~GHz in order to produce the best-fit to the pulse profile at 1.4~GHz. The dotted line shows the model of \citet{bcc+04} for a frequency of 1.4~GHz.}
	\label{scattercurve}
\end{figure}

In order to measure $\tau_s$ we used the pulse profiles at 6.5~GHz obtained from the survey and, where available, profiles at 1.4~GHz obtained elsewhere. This amounted to 12 pulsars in total. We assume that the scattering in the 6.5~GHz profile is negligible compared to that at 1.4~GHz and convolve this 6.5~GHz profile with an exponential scattering tail of decay time $\tau_s$, fitting against the 1.4~GHz profile. We ignored values of $\tau_s$ less than $2\,\mathrm{ms}$ where the results are not reliable, and further ignored any pulsars in which there was noticeable pulse profile evolution between the two frequencies. The values of $\tau_s$, and their standard deviation, for the remaining 10 are given in column 10 of Table~\ref{fluxpred}. Fig.~\ref{scattercurve} shows DM vs $\tau_s$ for this group of pulsars, with the relationship given by equation~(\ref{bhatscatter}) plotted.

Fig.~\ref{scattercurve} shows that the scattering tail size is clustered in the region from 10-100 ms, with the lower boundary imposed, as previously stated, by insufficient time resolution in the data to measure any small difference in profile shape at lower DMs. The higher boundary comes from the lack of known sources at $\mathrm{DM} > 1000\,\mathrm{cm}^{-3}\,\mathrm{pc}$, something this survey hoped to address when it began.
What it does show, however, is that high-DM ($\sim 1000\,\mathrm{cm}^{-3}\,\mathrm{pc}$) sources would be so highly scattered at 1.4 GHz that it is unlikely the PMPS would have discovered them, no matter how high their flux density. The fact that so few were found in this survey, indicates that the amount of obscuring material between ourselves and these putative sources, coupled with the lower flux density, prevented their discovery.
The four highest-DM sources in Fig.~\ref{scattercurve} also lie below the model of \citet{bcc+04}. Clearly, when observing at low frequencies, surveys are biased to detecting pulsars with less scattering, and thus the model of \citet{bcc+04} can be seen as a lower limit to scattering. However, our results suggest that the relation between scattering and DM may be flatter than previously thought.

\subsubsection{Simulated Population}\label{sec:simulation}
To fully understand the interplay between the spectral indices and degree of scattering, we produced a simulated population of pulsars in which both of these factors could be allowed to vary. Using the PSRPOP software\footnote{http://psrpop.sourceforge.net}, for each realisation, a simulated population of pulsars was produced with a different mean and standard deviation of the spectral index. All other variables were held fixed at the values given by \citet{lfl+06}, and are listed in Table~\ref{simpar}. The population was increased in size until it contained a subset of 1033 pulsars with flux density and position in the Galactic plane such that they would be detectable by the PMPS. As the PMPS is one of the most-studied pulsar surveys, responsible for the discovery of a large fraction of the pulsar population, it is suitable for constraining the population of pulsars in the Galactic plane. The survey parameters are outlined in Table~\ref{pmps_params}. PSRPOP models the population of `normal' pulsars, and so our simulation is unaffected by the acceleration effects which are most commonly observed for MSPs. The acceleration of a pulsar in its orbit can smear the pulse out such that it is not discovered in a blind search, if the data is not searched in acceleration space.

\begin{table}
	\begin{center}
	\caption{Parameters used to simulate the pulsar population used in Section \ref{sec:simulation}.}
		\begin{tabular}{lr}
		\toprule
		 Galactic latitude scale height & 330 pc \\
		 Minimum luminosity & $0.1\,\mathrm{mJy}\,\mathrm{kpc}^{2}$\\
		 Maximum luminosity & $10000\,\mathrm{mJy}\,\mathrm{kpc}^{2}$\\
		 $\mathrm{d}\log{N}/\mathrm{d}\log{L}$ & $-0.8$ \\
		 $\log\,$(Mean of P distribution (ms) ) & $2.4$\\
		 $\log\,$(Std. dev. of P distribution (ms) ) & $0.34$\\
		 \\
		 Number of pulsars detectable by PMPS & 1033\\
		\bottomrule
		\end{tabular}
		\label{simpar}
	\end{center}
\end{table}

\begin{figure*}
	\begin{center}	\includegraphics[width=17cm]{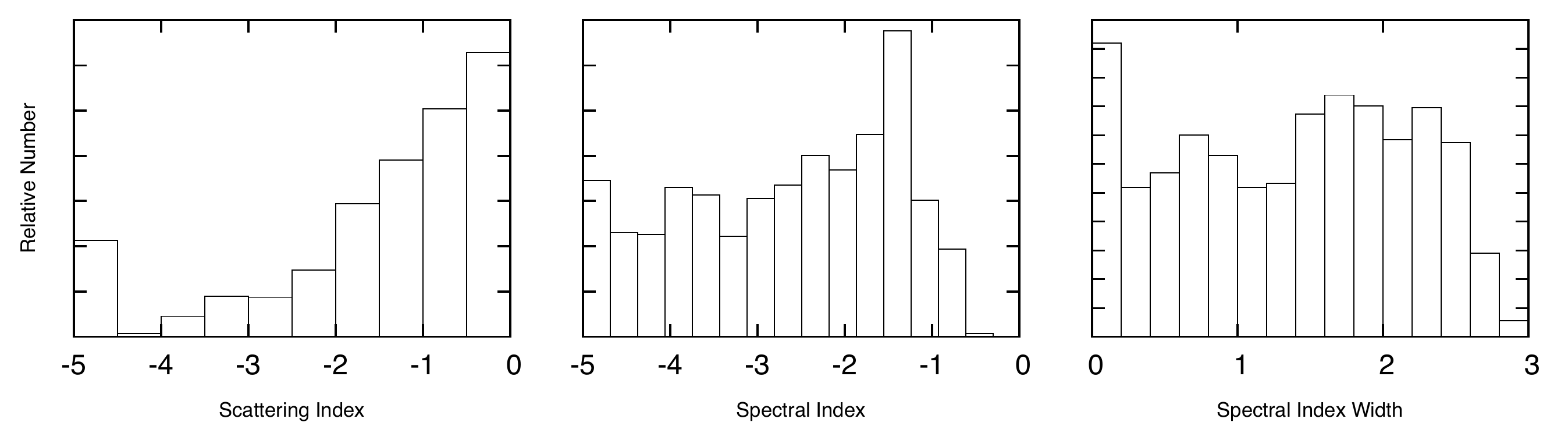}
	\end{center}
	\caption{Histograms showing the relative number of simulations (linear scale) consistent with our observations, for each of the three parameters.}
	\label{hist_plots}
\end{figure*}

\begin{figure*}
	\begin{center}	\includegraphics[width=17cm]{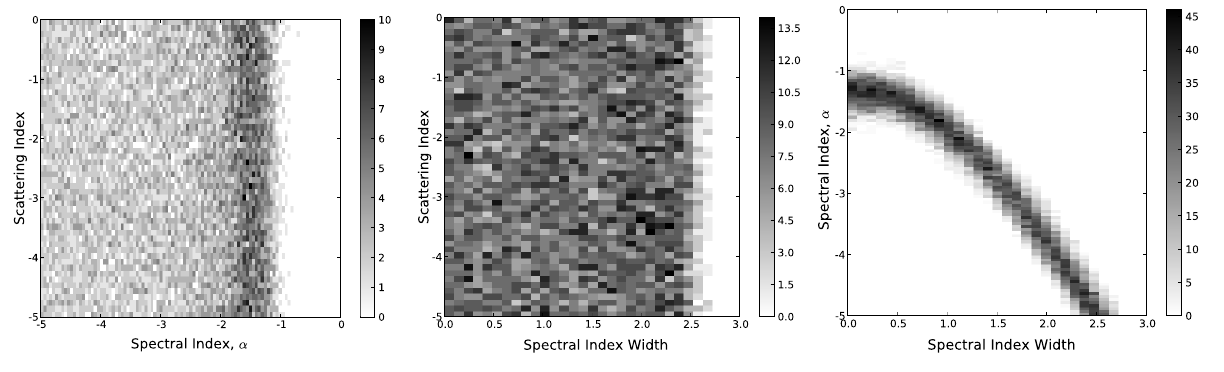}
	\end{center}
	\caption{Grey scale plots showing the number of simulations that were found to be consistent with our results, for different combinations of simulation parameters.}
	\label{sim_plots}
\end{figure*}

PSRPOP produces values of the flux density at 1.4~GHz, the pulse width and distance for each pulsar, based upon the analysis of the known pulsar population presented in \citet{lfl+06}. Using these, the flux density and pulse width were then recalculated for the entire simulated population at 6.5~GHz, each time assuming a different power-law dependance of the scattering tail size with frequency, from equation~(\ref{bhatscatter}) - labelled as ``scattering index" in Figs.~\ref{hist_plots} and \ref{sim_plots}. From these values, the number of pulsars which would be visible in the MMB survey region could be calculated for each simulated population.

\begin{table}
	\begin{center}
	\caption{Observational parameters for the Parkes Multibeam Pulsar Survey (PMPS)}
		\begin{tabular}{lr}
		\toprule
		%\multicolumn{2}{c}{Receiver Parameters} \\
		Number of beams  & 13 \\
		Polarizations/beam   & 2 \\
		Centre Frequency (MHz) & 1374\\
		Frequency channels  & 96 $\times$ 3~MHz \\
		System temperature (K) & 21 \\
%		\midrule
		\\
		%\multicolumn{2}{c}{Survey Parameters} \\
		Galactic longitude range & $260\degree$ to $50\degree$ \\
		Galactic latitude range & $|b| \leq 5\degree$ \\
		Sampling interval & 250 $\mathrm{\mu}$s \\
		Observation time/pointing & 2100~s \\
		Limiting Sensitivity for centre beam & 0.14~mJy\\
		\bottomrule
		\end{tabular}
		\label{pmps_params}
	\end{center}
\end{table}

We could then find all simulated populations in which the number of detected pulsars is within one standard deviation of the number that we detected. These results are plotted as histograms in Fig.~\ref{hist_plots} and on a greyscale in Fig.~\ref{sim_plots} for each pair of simulation parameters. From Fig.~\ref{sim_plots} we can exclude certain regions of the parameter space that are clearly inconsistent with our results.

For example, there is a clear favouring of spectral indices of $\sim-1.5$ in the histogram, as we might have expected given the results found in Section \ref{sec:SID}. It is, however, difficult to constrain the scattering index in any way, as some simulations fit our observations, regardless of the value it takes. 

We can also put quantitative limits on the number of pulsars with flat spectral indices, based on the small number of detections in our survey, by simulating a population with a fixed spectral index at $-0.5$, $0$ and $0.5$. By comparing the number of detections such a population would give compared to what we have observed, we can estimate a limit on how many such pulsars could be in the Galactic pulsar population. The simulations show that the population of pulsars with spectral index greater than $-0.5$ cannot exceed 10\% and cannot exceed 5\% for a spectral index of 0.

Our findings have relevance to future high-frequency surveys planned with the phase-1 Square Kilometer Array (SKA). Using the specifications as described in \citet{ska07}, we repeated this simulation for the case of 620 15-metre dishes with system temperature of 35~K and efficiency of 65\%, at an observing frequency of 10~GHz with a bandwidth of 500~MHz. In order to account for the apparent deviation of our results from those of \citet{bcc+04}, we obtained results using the scattering law described in equation~(\ref{bhatscatter}), and also using a linear fit to the data in Fig.~\ref{scattercurve}. 

Our simulation suggests that a survey of the inner Galaxy ($|l| \leq 45\degree$, $|b| \leq 5\degree$), using an integration time per survey observation of $1800\,\mathrm{s}$, could expect to find $\sim 2700$ pulsars, assuming a spectral index distribution with mean $-1.7$, standard deviation $0.3$ and a scattering index of $-3.8$. The predicted number of detections, $n$, changes by less than $\sqrt{n}$ whether using our linear fit, or that of \citet{bcc+04}, to model the dependance of scattering timescale with DM. However, we are yet to discover the true Galactic-centre pulsar population, and it is likely that when we do so, the degree by which this population is scattered will be much larger, requiring the use of high observing frequencies.

In order to compare the merit of such a survey to one at a lower frequency, a similar simulation was also performed for an observing frequency of 1.4~GHz. This survey would detect many more pulsars over same region, and would have a vastly improved survey speed. However, for Galactic latitudes $|b| \leq 2\degree$, the high-frequency survey is predicted to find three times as many pulsars with $\mathrm{DM}>900\,\mathrm{cm}^{-3}\,\mathrm{pc}$ as the low-frequency survey. Hence these two survey types are complementary to one another, provided the high-frequency survey is targeted at discovering the Galactic centre population of pulsars. This is not possible with a low-frequency survey of the same region, due to the large amount of scattering introduced by the material between ourselves and the Galactic centre.

\section{Conclusions}
A survey of the Galactic plane in the region $-60\degree \leq l \leq 30\degree, |b| \leq 0.25\degree$ was carried out between February 2006 and September 2007 using the 64-metre Parkes radio telescope at a frequency of 6.5~GHz. The seven beam Methanol Multibeam receiver was used to increase the amount of sky coverage in a single pointing, and reduce the length of time required to complete the survey. Three pulsars have been discovered from two passes through the survey data; PSR J1410$-$6132, a young pulsar with an extremely high DM of 960$\,\mathrm{cm}^{-3}\,\mathrm{pc}$, PSR J1834$-$0812, also with a high DM of 1023$\,\mathrm{cm}^{-3}\,\mathrm{pc}$, and PSR J1746$-$2850, close in angular distance to the Galactic centre and with a DM of $\sim 1000\,\mathrm{cm}^{-3}\,\mathrm{pc}$. This small number of detections indicates that any population of pulsars with flat spectral indices in the inner-Galaxy must be small.

Analysis of the known pulsars in the survey region has shown that the decrease in intrinsic flux density of pulsars as a result of the spectral index removes any benefit obtained from the reduced scattering at this frequency. Of the known pulsars seen in the survey data, the majority have a shallow spectral index, and the few that do not are intrinsically bright sources, which leaves them above our detection threshold, while one, J1550$-$5418, a magnetar, has emission with a negative spectral index, making it brighter at 6.5~GHz than at the typical survey frequencies of 1.4 GHz or lower. The date of this observation places an important limit on when the source first emitted radio pulsations, since \citet{gg2007} were unable to observe pulsations in August 2006.

The discovery of three pulsars with DMs in excess of $900\,\mathrm{cm}^{-3}\,\mathrm{pc}$ indicates that observing at higher frequencies does, indeed, allow one to probe at great distances in the direction of the Galactic centre. However, the reduction in dispersive delay makes distinguishing which sources are likely to be astrophysical extremely challenging. We do find, however, that there is some evidence that the relationship between scattering timescale and DM is flatter than suggested by the model of \citet{bcc+04}. This strengthens the case for future surveys at lower frequencies, for example a new 1.4 GHz multibeam survey at Parkes \citep{keith2010} and the all-sky survey using LOFAR \citep{bws}.

Future high-frequency surveys of the Galactic centre region with the SKA \citep{smits2008} could potentially discover thousands of Galactic-centre pulsars. However, this will require a large observing bandwidth to increase the dispersive delay across the band, and location of the telescope in a region with improved RFI conditions.

\section*{ACKNOWLEDGEMENTS}
We gratefully thank Bernd Klein, Kosmas Lazaridis and Ramesh Karuppusamy for their hard work and assistance with confirmation observations at Effelsberg.
S.B. gratefully acknowledges the support of STFC in his PhD studentship. D.R.L. and M.A.M. acknowledge support from a Research Challenge Grant provided by
the West Virginia EPSCoR program. M.A.M. is an Alfred P. Sloan research fellow.
The HYDRA supercomputer at the Jodrell Bank Centre for Astrophysics is supported by a grant from the UK Science and Technology Facilities Council. The Parkes Observatory is part of the Australia Telescope which
is funded by the Commonwealth of Australia for operation as a National
Facility managed by CSIRO.
\newpage

\bibliography{mmbrefs}
\bibliographystyle{mnras}

\label{lastpage}

\end{document}